\documentclass[12pt]{article}

\usepackage{graphics}
\usepackage{natbib}
\usepackage{indent}
\usepackage{rotating}

\bibpunct[;]{(}{)}{;}{a}{}{;}

\begin{document}

\baselineskip=0.9cm 


\begin{center}

{\large{\bf Impact of aerosols present in Titan's atmosphere on
the CASSINI radar experiment}}\\

\vspace{0.7cm}

S. Rodriguez$^1$, P. Paillou$^1$, M. Dobrijevic$^1$, G.
Ruffi\'e$^2$, P. Coll$^3$, J.M. Bernard$^3$, P. Encrenaz$^4$\\
\vspace{0.7cm}

E-mail: rodriguez@obs.u-bordeaux1.fr\\
\vspace{1.5cm}

$^1$OASU-L3AB UMR 5804, BP 89, 33270
    Floirac, France\\
$^2$ENSCPB-PIOM, Universit\'e Bordeaux I, 33400 Talence, France\\
$^3$LISA, UMR 7583, Universit\'e Paris 12, 94010 Cr\'eteil, France\\
$^4$LERMA, UMR 8540, Observatoire de Paris, 75014 Paris, France\\

\vspace{2cm}

Submitted to Icarus (journal)\\
Submitted 09/10/2002 (date)\\
Revised ----------- (date)\\

\vspace{2cm}

Manuscript Pages: 43\\
Tables: 5\\
Figures: 7\\

\end{center}


\clearpage

\noindent{{\bf Proposed Running Head:} Impact of Titan's
atmosphere on the Cassini radar}\\
\vspace{1.5cm}

\noindent{{\bf Editorial correspondence to:}} \\
S\'ebastien Rodriguez \\
Laboratoire d'Astrodynamique, d'Astrophysique et d'A\'eronomie de
Bordeaux \\
2, rue de l'Observatoire \\
B.P. 89\\
33270 Floirac, France. \\
Phone: 33-5-57-77-61-19 \\
Fax: 33-5-57-77-61-55\\
E-mail: rodriguez@obs.u-bordeaux1.fr\\


\clearpage

\noindent{{\bf ABSTRACT}}

Simulations of Titan's atmospheric transmission and surface reflectivity
have been developed in order
to estimate how Titan's atmosphere and surface properties could affect
performances of the Cassini radar experiment.  
In this paper we present a selection of models for Titan's haze, vertical rain distribution,
and surface composition implemented in our simulations.
We collected dielectric constant
values for the Cassini radar wavelength ($\sim\:2.2$ cm) for materials of interest for 
Titan: liquid methane, liquid mixture of methane-ethane, 
water ice and light hydrocarbon ices. Due to the lack of permittivity values for
Titan's haze particles in the microwave range, we 
performed dielectric constant ($\varepsilon_r$) measurements around $2.2$ cm on tholins 
synthesized in laboratory. We obtained a real part of $\varepsilon_r$
in the range of 2-2.5 and a loss tangent between $10^{-3}$ and $5.10^{-2}$.
By combining aerosol distribution models (with hypothetical condensation at 
low altitudes) to surface models, we find the following 
results: (1) Aerosol-only atmospheres should cause no loss and are 
essentially transparent for Cassini radar, as expected
by former analysis. (2) However, if clouds are present, some atmospheric models 
generate significant attenuation that can reach $-50\:dB$,
well below the sensitivity threshold of the receiver. In such cases, 
a $13.78\:GHz$ radar would not be able to measure echoes 
coming from the surface. We thus warn about possible risks of misinterpretation
if a \textquotedblleft wet atmosphere\textquotedblright $\:$is not taken into account.
(3) Rough surface scattering leads to a typical response of $\sim\:-17\:dB$. These results 
will have important implications on future Cassini radar data analysis.

\vspace{1cm}



\clearpage

\section{Introduction}

Titan is the second largest satellite in the solar system. It is
the only one known to possess an extended and dense atmosphere,
hiding its surface. Primarily composed of N$_2$ with less than
10\% CH$_4$ and 1\% H$_2$, Titan's thick atmosphere is subject to
an active chemistry induced by solar insolation, solar wind and
charged particles from the saturnian magnetosphere. This chemistry
results in the production of a great
diversity of hydrocarbons and nitriles in the stratosphere with increasing complexity
and molecular weight
\citep{1984yung,1995toublanc,1996lara,1998lara}. Settling in 
the atmosphere, chemistry end products form an extensive system of
organic aerosol haze responsible for Titan's characteristic orange
color and hiding the lower atmosphere and the surface at
UV-visible-NIR wavelengths. It is also possible that lighter
hydrocarbons or nitriles diffuse downwards and 
condense as liquids or solids on aerosols below altitudes of 85 km.

Ground-based observations, as well as Pioneer and Voyager
missions, have brought a great number of constraints on the haze
properties but revealed the difficulty to precisely
determine not only its horizontal and vertical structure but also
its chemical composition. Concerning the
surface, radar observations and microwave radiometry or near-infrared
studies only reveal the presence of a heterogeneous
surface plausibly composed of randomly distributed patches of
water ice and organic compounds, solid or liquid
\citep{1990muhleman,1995coustenis,1997coustenis,2001coustenis}.
The precise nature of Titan's surface is still completely unknown.
With the hope of disclosing the remaining questions about its composition, the CASSINI-HUYGENS spacecraft - 
a joint NASA/ESA/ASI mission - was launched in 1997 and will reach Saturn in 2004. 
The spacecraft, composed of an orbiter
and a probe, will favour the exploration of Titan. A
multipurpose radar was included in the scientific payload of the
Cassini orbiter. Three active modes (SAR - for Synthetic Aperture Radar -, altimeter and scatterometer) and one
passive mode (radiometer) will be used within
the Radar experiment. Thanks to its active modes, the Radar instrument
is expected to be able to observe Titan's surface through its
dense atmosphere and thus determine whether oceans are present
and map the geology and the topography of the solid surface of Titan.

These scientific goals rely on
analysis of the received radar echoes and the analysis depends on models
describing microwave transmission in the atmosphere and on surface
reflectivity. In this paper, we present a
quantitative study of the possible impact of particles present in
Titan's atmosphere on the Cassini Radar experiment and
some results about surface reflectivity. We first present the models
of Titan's surface and atmosphere we selected as a basis for our computations.
Then, in section \ref{exp} we show results of experiments designed to determine the
electromagnetic properties of titanian aerosol analogs - called
tholins. We finally discuss in section \ref{simul} the possible
behaviour of the Cassini Radar in presence of atmospheric haze and rain on the basis of
simple simulation of radar wave propagation for various models for aerosol and rain particles
formation. We also show
some simulations of the backscattered power by smooth and rough
surfaces covered with tholins.

\section{Distribution of condensates in Titan's atmosphere and on its surface}
\label{distributions}

\subsection{Meso- and stratosphere: vertical distribution of aerosol haze}
\label{haze}

Since 1973, there is observational evidence for the existence
of an aerosol component in Titan's upper atmosphere. 
Particles with size greater than molecules
were invoked to reconcile models of Titan's upper atmosphere 
and polarimetric and photometric observations \citep{1973veverka,1973zellner,1980rages}. Several 
radiative transfer models incorporating aerosols were published,
considering the shape and optical
properties of these aerosols as free parameters, as well as their vertical
distribution in terms of size and number density. These
models were regularly adjusted to be consistent with the growing
number of observational data of increasing resolution.
First generation models \citep{1973danielson,1977podolak} assumed
a homogeneous atmosphere where dust particles and gas were uniformly
mixed above the surface of Titan. \citet{1979podolaka}, \citet{1980rages}, 
\citet{1980tomasko}, and \citet{1983rages&pollack} used in their calculations 
multi-layer models and \citet{1982tomasko} used the first totally 
inhomogeneous models with a gradient in particle size with the optical depth.
They constrained the aerosol mean radius to between $0.1\:\mu m$ and $0.5\:\mu m$,
but always failed to simultaneously fit the polarimetric and photometric data 
acquired by Pioneer and Voyager missions when considering spherical 
particles of single size mode. 

Further studies greatly improved the model fit to data with the use of more complex 
haze distribution models. Inspired by the pioneer work of \citet{1980toon}, 
\citet{1989mckay} considered Titan's aerosols properties in the new light of 
microphysical modeling, with the aim of integrating these results in a 
radiative-convective model of the thermal structure of Titan's atmosphere. 
They assumed that all particles at a fixed altitude had the same radius and 
that only thermal coagulation, sedimentation, and electrostatic forces are 
responsible for variations in particle size and concentration with altitude. 
They predicted a size of $\sim\:0.4\:\mu m$ for the
aerosols around an altitude of 85 km, in good agreement with previous values obtained from
observational fits. In the same way, \citet{1992cabane} developed
a model including sedimentation and eddy diffusion
accounting for transport processes, gravitational (coalescence)
and thermal coagulation for collection processes.
Their calculations focused on the prediction of the
distribution and radius of aerosol particles throughout the
atmosphere between the higher formation region and the lower stratosphere
(above 85 km), before condensation of hydrocarbons and nitriles
on the aerosols could occur. The final particle radius obtained 
in the low stratosphere is about $0.2\:\mu m$. 
In the meantime, \citet{1990frere} pursued the development of a
complete microphysical modeling of Titan's aerosols, including for
the first time condensation processes of organics and nitriles
present in the gas phase in the low stratosphere. Above 100 km, the profiles computed by  \citet{1989mckay},
\citet{1990frere}, and \citet{1992cabane} are quite similar.
The recent microphysical modeling taking into account aerosol 
growth into fractal aggregates \citep{1993rannou,1993cabane,1995rannou} gives
concentrations and effective profiles (Fig. 1 and 3 in
\citet{1993cabane}) close to those given by classical
microphysics dealing with spherical particles (close enough for the precision in radius and concentration we need),
so we will not consider this model here. Besides in the wavelengths range we consider, fractal nature of 
micron-sized scatterers would not have significant effect on scattering and absorption calculations.
Figure \ref{heterogmodels}-(a) shows profiles of aerosol radius and 
concentration versus altitude resulting from computation by \citet{1989mckay}, 
\citet{1990frere} and \citet{1992cabane}.

\subsection{Low stratosphere and troposphere: models for clouds and rain}

After a rapid lapse rate below 120 km altitude, the thermal
profile of Titan reaches a minimum of 74 K at the tropopause (42
km). This thermodynamical cold trap region certainly induces
saturation (even supersaturation for the dominant and more
volatile species) and probably gas/liquid or solid phase
transitions. These transitions should occur for most of the
organics and nitriles present in gas phase if seed nuclei are
present \citep{1984saganb,1990frere}. If nucleation can be
activated, the most relevant scenario is that individual aerosol
particles may act as condensation nuclei, accrete hydrocarbons and
nitriles with more or less efficiency depending on 
solubility properties of the aerosols, and rain out when they pass through
saturated regions of the troposphere
\citep{1993lorenza,2001mckay}. However, observational evidence for a dense, 
persistent and ubiquitous cloud coverage in the supersaturated regions is not 
yet conclusive. \citet{1998griffith,2000griffith} hemispheric
integrated observations of Titan suggested a cloud cover. Titan's large cloud system
that might have been observed in September 1995 \citep{1998griffith} 
displayed cloud tops at an altitude of $z\:\sim\:15\:\pm\:10\:km$, covering about
10\% of Titan's disk, and seemed to live for
more than 2 days. Analysis of more recent spectra \citep{2000griffith}
suggested another identification of tropospheric clouds. It
revealed that clouds should reside at the altitude of
$z\:\sim\:27\:km$. These clouds covered $\sim\:0.5\:\pm\:1$\% of the 
moon's disk, and dissipated in only 2 hours. Thus, if clouds exist, 
they seem to be patchy, occupying only a fraction of Titan's disc. It may 
also be an intense but transient and fickle phenomena, making them even
more difficult to observe and firmly identify.

Several models treating condensation in
Titan's saturation region have been developed 
\citep{1988toonb,1990frere,1995courtin,1997samuelson&mayo}, aiming at the determination of cloud
properties. \citet{1988toonb} and \citet{1995courtin} studied the
characteristics of hypothetical pure methane clouds, considering 
atmospheric radiative properties. In
particular, \citet{1988toonb} assumed that if the mixing ratio of
methane exceedzd 2\% by volume near and below the tropopause, it
would reach (and even exceed) saturation level, triggering its
probable condensation. Re-examining the
Voyager IRIS infrared spectra of Titan, \citet{1988toonb} obtained an
optimal fit when adding a methane cloud composed of droplets with
a radius between $50\:\mu m$ and $3\:mm$ and a density between $1$
and $3500\:m^{-3}$. This cloud cover, that can be considered as an
ubiquitous homogeneous layer in addition to the aerosol haze,
should extend from 10 km  to 30 km. The new re-interpretation of IRIS
spectra led \citet{1995courtin} to quite different conclusions. Their
results, in best agreement with the data, did not contain a cloud
layer. They found however that if a condensed $CH_4$ cloud is
added, it must have a mean particles radius of about $50-60\:\mu m$ and
be located near the tropopause (about 40 km in altitude), though they
could not put constraints on its vertical extent. A combination of cloud
characteristics inferred from \citet{1988toonb} and
\citet{1995courtin} are summarized in Table 1.

{\bf [Table 1]}

\citet{1990frere} was the first to present a microphysical approach for 
condensation processes in Titan's low atmosphere, resulting in an estimation 
of the vertical radius and density distribution of aerosols. They elaborated a
model dedicated to the determination of the distribution and
chemical composition of the particles in the low atmosphere by
taking into account differential condensation of several atmospheric compounds 
on aerosols in the cold trap region, in addition to classical transportation
and collection processes. In their model condensation nuclei swell up, 
progressively covered with concentric layers of volatile compounds
as soon as they reach saturation; supersaturation is not allowed. 
They constrained the size of particles by the volume of material
available for condensation. They found that no significant
condensation occurs above 100 km. The gas to liquid or solid phase transition
starts modifying the particle structure below 90 km. 
From $0.1$ to $1\:\mu m$ around 100 km, the particle mean radius
increases rapidly due to sustained accretion and reaches a few tens
of microns at 85 km.  At this stage of the condensation process, the major
constituents surrounding aerosols are nitriles, mainly $HCN$ and $C_4N_2$ 
in solid state. A growing drizzle takes place down to the tropopause. 
Below 70 km, condensation of propane ($C_3H_8$) becomes predominant, 
then come $C_2$ hydrocarbons (with decreasing altitude: $C_2H_2$, 
$C_2H_6$ and $C_2H_4$). Between 30 km and the surface, the particle radius 
becomes greater than $200\:\mu m$, reaching a maximum of $\sim\:900\:\mu m$ at 11 km.
At these altitudes, liquid methane is the main constituent that condenses.
Below 11 km, the strong positive temperature gradient triggers methane layer
evaporation of the droplets. The fusion/sublimation rate is moderate
until 2 km above the surface. Then, the process becomes very efficient and
removes almost all the liquid around the aerosols. The particle
radius decreases to about $50-100\:\mu m$. That gives rise to a mist of 
ethane-enriched aerosols just above the surface (a \textquotedblleft
ghost\textquotedblright $\:$ of the raincloud, consistent with considerations by
\citet{1993lorenza}). Variations of the droplet
density follow those of the radius but in the opposite way 
due to collection processes. More recent studies of rain occurence 
in Titan's lower atmosphere \citep{1997samuelson&mayo} led to similar
conclusions as those of \citet{1990frere}. They obtained the same altitude
for the maximum particle size ($z\sim11-12\:km$) and roughly the same size profiles 
versus altitude.  The experimental work of \citet{1996mckay}
and recently \citet{1999colla} argued in favour of 
droplet chemical structure and size proposed by \citet{1990frere} (Fig. \ref{heterogmodels}-(b)).
\citet{1996mckay} and \citet{1999colla} demonstrated the insolubility 
of aerosol material in hydrocarbons and their relative solubility in nitriles. 
It is also consistent with the identification of nitriles 
in solid phase in Voyager IRIS spectra ($C_4N_2$ \citep{1987khanna,1999coustenis} and $HCN$,
$HC_3N$, $CH_3CH_2CN$ \citep{1999coustenis} that could constitute a 
$10\:\mu m$-hail around the altitude of 90 km).  
Aerosols might need a first nitrile layer to activate hydrocarbon 
condensation, as predicted by \citet{1990frere} (Fig. \ref{heterogmodels}-(b)). 

{\bf [Figure 1]}

\subsection{Surface}
\label{surface}

Very little is known about the surface of Titan, which is hidden
from view by the organic haze. Voyager 1 and 2 flybys allowed some
progress in the knowledge of the puzzling nature of the
satellite's surface, though they failed to directly reveal surface
details. Surface properties were only indirectly
inferred from the atmosphere characterization performed by the
spacecrafts. Voyager 1 and 2 flybys allowed the determination of
the atmospheric properties with an accuracy such that it was
possible to propose some new and more relevant models for Titan's
surface nature and to postulate the existence of a deep,
global ocean of hydrocarbon \citep{1983lunine}. The fact that
methane is likely to be depleted in the stratosphere within barely
10 million years due to a photolytical sink, coupled with the fact
that Titan's surface temperature is close to the triple point of
methane, supports the idea that Titan should possess a liquid
hydrocarbon reservoir at its surface or at least in its subsurface
\citep{1983eshleman}.

Observations have become available which suggest that Titan
is not covered by an ubiquitous ocean. Subsequent ground-based
radar observations, passive microwave radiometry and near-infrared
observations give us today a new insight in the current
understanding of Titan's surface. Direct radar sensing of Titan's
surface at a wavelength of $3.5$ cm
\citep{1990muhleman} revealed a surprisingly high reflectivity
(much higher than the reflectivity of a global methane-ethane
ocean) but still lower than that of the icy Galilean satellites.
Radar reflectivity was also extremely variable from one night to
the other, suggesting a heterogeneous surface. Passive radiometry
undertaken at the same time by \citet{1990muhleman} suggested an
emissivity of about 0.9 at similar wavelengths (higher than
those of Europa and Ganymede, but similar to that of Callisto).
The only way they found to reconcile high emissivity and high
reflectivity was to consider a surface covered with
\textquotedblleft dirty\textquotedblright $\:$ ice (ice
contaminated by organic compounds). This hypothesis was recently
confirmed with the help of near-infrared observations from HST (Hubble Space Telescope),
CFH (Canada-France-Hawaii) and, AEOS (Advanced Electro Optical System) telescopes in 
the methane-window regions ($0.94$,
$1.08$, $1.28$, $1.58$, $2$ and $5\:\mu m$), where it may be possible to probe the surface
between methane absorption bands.
These observations are consistent with dirty ice, including areas
of hydrocarbon seas \citep{1993lorenzb,1997lorenz}.
\citet{2001coustenis} have investigated the nature of ice present
at the surface of Titan. They took images (albedo maps) of Titan's
surface at $1.3$ and $1.6\:\mu m$ with a resolution high enough to
resolve Titan's disk with 20 pixels. They observed the same bright
feature at the equator that was previously observed by the HST \citep{1996smith} and the
Keck I telescope \citep{1999gibbard}. That allowed them to propose some
candidates spectrally relevant for the icy composition of the
different (bright/dark) regions observed. By comparison to other
surface spectra \citep{1995coustenis,1997coustenis}, they
suggested the presence of methane (or ethane) frost, ammonia ice,
and water ice on Titan's surface, probably covered by sediments of
organics, as foreseen by photochemical atmospheric models.
\citet{1980toon} predicted indeed that the aerosols produced in
the upper stratosphere should settle and deposit on the ground at
the rate of $0.1\:\mu m$ each year, to form a layer of about 400 m
over geological times. The hydrocarbon and nitrile liquids they
drag towards the surface are expected to stream down the relief
and fill lakes or seas, whereas solid organic cores of the
droplets should form sedimentary layers. It is nevertheless
impossible to discriminate whether aerosols are 
intimately mixed with ice or whether they constitute a dark
organic layer over an icy surface with some bright regions of
exposed ice. The aerosols falling over a lake or a sea do not sink
and create an organic layer, due to their poor solubility into
hydrocarbon liquids.

\section{Electromagnetic characterization of titanian materials}
\label{exp}

\subsection{Electromagnetic characterization of tholins}

Electric properties of titanian aerosol analogs were often
measured in visible-UV range, but are poorly known in the
microwave range. The only published value for dielectric constant
at GHz frequencies is the one taken from \citet{1984khare}
measurements. They found a permittivity of $4.71 - j0.013$ at
$f\sim300\:\:GHz$. Considering that tholins are major compounds of
Titan's atmosphere susceptible to interact with Cassini radar electromagnetic
wave, we performed a set of measurements of aerosols' analogs
permittivity in the GHz frequency range.

\subsubsection{Synthesis of tholins}

Since the early work of \citet{1973khare}, many laboratory
simulations of Titan's atmosphere have been achieved. Such
experiments were conducted by irradiation with UV light,
electrical discharge, or energetic electrons, of a gas mixture
close to the Titan's atmosphere composition, isolated from
contamination in a glass recipient -- the reactor. Without any
exceptions, these simulations produced a solid organic material --
termed tholin -- which has optical properties similar to those
needed to match the geometric albedo of Titan
\citep{1980rages,1989mckay}.

In order to get tholin samples for further dielectric
measurements, we conducted experiments in the LISA laboratory
where a reactor essentially dedicated to the study of
the titanian atmospheric chemistry was designed. A detailed description of the experimental 
devices can be found in \citet{1999colla} and \citet{1999collb}. According to
\citet{2001mckay}, this experiment promises to give the most
accurate simulation of Titan's haze formation, as the dinitrile $C_4N_2$ 
(detected in solid phase on Titan), has been identified for the first time as 
a product of the simulation, this clearly constitutes a validation of the 
experiment \citep{1999collb}.

In our case, we performed three titanian atmosphere simulations with
a low pressure flux ($\sim\: 1-4\:mbar$ of total pressure) of a $N_2/CH_4$ 
mixture (in a 98:2 proportion). 
A glass tablet or a silica cylinder was inserted into 
the reactor where tholins are deposited to form a film. In order to
collect tholin samples, the reactor was installed inside a glove
box filled with nitrogen to avoid contamination of the
sample by the atmosphere in the room, in particular oxygen atoms, before
electromagnetic characterization. Finally, the volume of deposited tholins
- an essential parameter for future permittivity measurements - is
determined using scanning electron microscopy to measure the
thickness of the sample. Table 2 summarizes synthesis conditions and 
volume estimations for each set of experiments.

[{\bf Table 2}]

\subsubsection{Dielectric constant measurements}
\label{tholineps}

Tholin samples produced in the LISA reactor were characterized
at the PIOM laboratory in order to measure their dielectric
properties in the GHz range. Permittivity measurements were performed
using a rectangular resonant cavity pierced on its top with a hole
used to insert the sample. For the purpose of these experiments,
we used two different kind of cavities : one at 10 GHz (i.e.
$\lambda=3\:cm$) and a second one at
2.45 GHz (i.e. $\lambda=12.2\:cm$). As
the change in the dielectric constant value should be small
in the range 1-10 GHz, the choice of a cavity was made only
because of the sample holder shape : the 10 GHz cavity is too
small to allow a large rectangular hole on its top center and only
a cylindrical sample holder can be used. Each cavity is composed of
a wave-guide short-circuited at each end. Excitation and
detection of resonance inside the cavity are done by co-axial
transitions mounted on connectors linked to a network 
analyzer (ANRITSU 37325A).

The resonance modes that can be established in a cavity are
deduced from the theory of wave propagation along a wave-guide.
These modes are characterized by three integer indices $m$, $n$
and $p$. In the case of a rectangular cavity, $TE01p$
resonance modes are usually used ($m=0$ and $n=1$ as dominant
mode, $p$ representing the number of maxima of the electric field
along the wave-guide length). In particular, the index $p$
is chosen odd ($p=3$) in order to ensure the position of a maximum
in the center of the cavity, where we insert the sample. The
sample holder symmetry axes are set in a way that they are parallel
to the electric field established in the cavity, in order to
minimize perturbations due to depolarization effects.

Once the cavity is connected to the network analyzer, we first introduce
an empty sample holder into the cavity to measure the resonance
frequency $F_0$, and the quality coefficient $Q_0$ of the cavity as
blank references. Then, we insert an identical sample holder --
same glass/silica volume -- but covered with a tholin
film. The introduction of a tholin sample triggers a deviation
from the resonance frequency, $\Delta F$, and an additional
attenuation conducting to a new quality factor $Q_1$. 
As it is well established for resonant cavity experiments, the measurement of $\Delta F$,
$Q_0$, and $Q_1$ allows to directly obtain the dielectric constant
$\varepsilon^{'} - j\varepsilon^{''}$ of the sample thanks to the relations \citep{1971boudouris}:

\begin{equation}
\label{equaF}
\frac{\Delta F}{F_0} = - \alpha(\varepsilon^{'} - 1)
\end{equation}

\begin{equation}
\label{equaQ}
\Delta(\frac{1}{Q_1}-\frac{1}{Q_0}) = 2 \alpha
\varepsilon^{''}
\end{equation}

with $\alpha = \alpha_0 \times V_{sample}$, $V_{sample}$ being the
volume of the tholin sample and $\alpha_0$ being the cavity filling
coefficient (ratio between the tholin sample + sample holder volume and
the cavity volume).

For each set of tholin samples (see Table 2), we obtained a value for the tholin 
dielectric constant. Results of our measurements are shown in Table 3. 
While real part of the dielectric constant shows a very little variation 
($\sim\:10$\%), the imaginary part varies by a factor of 50. 
This variation is due to the poor estimation of 
factor $Q$, which is highly sensitive to the volume of material introduced in the 
cavity. An accurate value for this volume is very hard to obtain.

{\bf [Table 3]}

\subsection{Electromagnetic characterization of other titanian materials in liquid and solid states}

Apart from tholin characterization, we also searched in the
literature for values of permittivities of
materials of interest for radar sounding of Titan's atmosphere and
surface, in the microwave range.

\subsubsection{Pure liquid methane}

The dielectric properties of pure liquid methane and their
implications for Titan's study were investigated by
\citet{1990thompson&squyres}. They published a rough estimate of
$\varepsilon_r$ for liquid $CH_4$ around $94\:K$:
$\varepsilon^{'}\sim1.7$ and
$\varepsilon^{''}\sim1.5\times10^{-2}$. This is fully consistent
with the value found in {\it Handbook of Chemistry and Physics}
(1988-1989) around $100\:K$ in the hundreds of MHz range.

\subsubsection{Titan's raindrops and hypothetical surface oceans: liquid hydrocarbon and nitrile mixtures}

According to \citet{1990frere}, Titan's raindrops are primarily
made of light hydrocarbon liquids instead of pure methane,
progressively accreted in concentric layers around aerosols.
During the fall of the raindrops, internal dynamics stirs the
liquid $CH_4$ and $C_2H_6$ into a homogeneous mixture. Such a
mixture was the subject of a few experimental studies aiming at
determining its dielectric properties. \citet{1990muhleman} indicate
that the real part of the dielectric constant for liquid
hydrocarbons is well known to be in the range of 1.6 to 2 and that
loss tangents are of the order of $10^{-3}-10^{-4}$ in the GHz range.
\citet{1979singh} found a real part of permittivity between 1.77
and 1.9 for methane-ethane mixtures at low temperature (typically
around 100 K) and \citet{1992ulamec} obtained values between 1.71 and 1.95 around
94 K. The values from \citet{1979singh} were obtained
using a fit to an equation derived from the Clausius-Mossotti
function, on the basis of liquid compressibility considerations,
and those coming from \citet{1992ulamec} were inferred from
capacity measurements. We have no indications about what could be
frequency range they considered. However, such values are consistent
with dielectric constant measurements of \citet{1992sen} for
liquid alkanes and hydrocarbon mixtures at 1.2 GHz and 298 K.

\subsubsection{Titan's surface: hydrocarbons and water
ice}

Titan's surface is supposed
to be composed of various icy materials ($H_2O$ ice, hydrocarbons
in solid state) due to its low 94 K temperature \citep{1990lellouch},
probably fractionally covered or mixed with
tholin material. It then should present a high and contrasted
reflectivity in the radar wavelength range. Dielectric constants of such 
materials were experimentally measured. We had access to
precise values for the real part of their permittivity, but only
to rough estimations for the imaginary part: $H_2O$ ice corresponds to
$\varepsilon^{'}\:\sim\:3.1$, measured around 10
GHz and valid within the temperature range from -1 to -185$^o$C
(\citet{1982ulaby}, vol. 3, appendix E, p. 2028), and a
$CH_4-C_2H_6$ ice mixture has $\varepsilon^{'}\:\sim\:2-2.4$ in the
same wavelength range \citep{1990thompson&squyres}.

All the dielectric constants of interest for radar sounding of
Titan are summarized in Table 4.

{\bf [Table 4]}

\section{Simulation of the CASSINI radar behaviour}
\label{simul}

\subsection{Testing the radar penetration through the atmosphere}

In this part, we present simulations of the CASSINI Radar
experiment in active modes. This section mainly deals with the 
dielectric models we have developed to quantify the impact of condensates
in Titan's atmosphere. 

We used in our calculations the complex dielectric constant we 
measured for tholin material. Uncertainties 
on the permittivity of tholins have repercussions on the attenuation we calculate: 
concerning atmospheric transmission simulations, the attenuation 
does not vary linearly with the dielectric constant. 
The minimum value for attenuation was obtained with $\varepsilon^{'} = 2.1$ and 
$\varepsilon^{''} = 0$, and the maximum attenuation corresponds to $\varepsilon^{'} = 2.1$ and
$\varepsilon^{''} = 0.1$. For the preliminary results on homogeneous atmosphere 
models, we considered the mean value measured for the dielectric constant of tholins: 
$\varepsilon_{r}=2.2-j0.05$. 

\subsubsection{Description of the Rayleigh-Mie scattering model}

We consider the propagation of a radar wave through Titan's atmosphere.
The simple model we developed enables us to determine the
attenuation an incident radar wave (emitted power $P^{inc}$) will
undergo while crossing the dense atmosphere of Titan. The
attenuation in decibels is given by

\begin{equation} Att_{total}(dB) = Att_{gases} +
Att_{particles} = 10\:log_{10} (\frac{P^{ground}}{P^{inc}})
\end{equation}

with $P^{ground}$ being the radar wave power reaching the
surface of Titan.

The gaseous component is considered with no ambiguity as totally
transparent for the radar and its permittivity is set equal to one. Thus,

\begin{equation}
Att_{gases}(dB) = 0
\end{equation}

Only the condensate -- aerosols, hail and rain particles -- component could
have sufficient size, density and permittivity to interfere with
the radar wave. In this case, we can reasonably simulate the medium that the
wave will encounter before reaching the surface by a diluted
medium composed of dielectric spheres in suspension in vacuum.
Only single scattering and absorption of the incident wave are
considered here and we can then write for an homogeneous atmosphere:

\begin{equation}
\label{bouguer} \frac{P^{ground}}{P^{inc}} = e^{-[C(\sigma_{scat}
+ \sigma_{abs})\frac{h}{\cos{\theta}})]}
\end{equation}

 with,

\[
\begin{array}{lp{0.8\linewidth}}
    C,              & particle density, \\
    h,              & thickness of the atmosphere, \\
    \theta,         & incidence angle of radar wave (set to $0^{o}$ for the altimetric mode),\\
    \sigma_{scat},  & single scattering cross section, \\
    \sigma_{abs},   & absorption cross section.
\end{array}
\]

Cross sections for spherical scatterers are taken from Rayleigh
and Mie theories. Transition between the Rayleigh and Mie regimes
occurs when the particle radius is greater than $r_{lim}=0.05 \time \lambda$.
Rayleigh formalism is valid only for small particles in comparison
to the wavelength. In the case of the Cassini Radar, 
the upper limit to the radius $r_{lim}$ of the scatterer is set to
$1.08$ mm.

Rayleigh scattering and absorption cross sections can be
analytically solved and are given by \citet{1997ishimarua}:

\begin{equation}
\sigma_{scat} = \frac{128\pi^5r^6}{3\lambda^4}|\frac{\varepsilon_r
- 1}{\varepsilon_r + 2}|^2 \\
\end{equation}

\begin{equation}
\sigma_{abs} =
\frac{8\pi^2r^3\varepsilon^{''}}{3\lambda}|\frac{3}{\varepsilon_r
+ 2}|^2
\end{equation}

Mie scattering cross sections are numerically computed by means of
a classic BHMIE code taken from \citet{1983bohren}. The complex
parameter $\varepsilon_r = \varepsilon^{'} - j\varepsilon^{''}$
represents the dielectric constant of the spheres of radius $r$.
It should be frequency-dependent and was experimentally measured
around the Cassini Radar experiment frequency (cf. section \ref{exp}).

{\bf [Figure 2]}

In our model (for heterogeneous atmosphere cases), the atmosphere of Titan is vertically cut off into
homogeneous layers containing particles of fixed size and density
(Fig. \ref{model}). We then
compute the ratio between outgoing and incident flux
$P^{out}_i/P^{inc}_i$ for each layer $i$ of thickness $h_i$. The
resulting final attenuation is the sum of attenuations induced by
each layer $i$. It is given by :

\begin{equation}
Att_{total}(dB) = \sum_{i}Att_i = \sum_{i}10\:log_{10}
(\frac{P^{out}_i}{P^{inc}_i})
\end{equation}

Equation (\ref{bouguer}) indicates that the part of the flux
scattered or absorbed by the particles is lost. It reproduces well
what could happen to a wave in single scattering regime along the
radar line-of-sight.

\subsubsection{Homogeneous atmospheres}

We computed the attenuation of the centimetric signal emitted by the
Cassini Radar instrument induced by the presence of particles in a
ideal Titan-like homogeneous atmosphere. In this simple case, we
considered single crossing of the wave through a totally
homogeneous atmosphere, made of a unique layer with a single
mode of particle size, density and dielectric properties.

The thickness $H$ of the homogeneous layer was constrained
by the aerosol haze extension within Titan's atmosphere and was
fixed to $h=500\:km$. We fixed $\varepsilon_r$ to its experimental mean
value of $2.2-j0.05$. The only free parameters in our simulation
were particle radius and density. We decided to put a constraint on 
these parameters from extreme values found in
literature: density ranges between unity and a maximum
around $10000\:cm^{-3}$, and radius ranges between a few nanometers and a
high value of $20\:\mu m$. Attenuations induced by such
particles over a 500 km thick
layer are shown in Fig. \ref{homorealistic}-(a). Each point of the
figure represents the attenuation calculated for a unique
atmospheric scenario, {\it i.e} a unique pair of particle radius and
density over a homogeneous Titan atmosphere.

{\bf [Figure 3]}

The detection threshold (noise equivalent $\sigma_0$) of the
Cassini Radar instrument is limited to about $-25\:dB$ \citep{1991ieee}.
It means that, with the hypothesis that all the power reaching the surface
of Titan is backscattered to the radar, a signal attenuation greater than 12.5 dB
for the first crossing of the atmosphere would lead to no signal detection by the instrument.
Figure \ref{homorealistic}-(b) is a
2-D representation of the variation of the attenuation versus
particle radius and density. It shows the position of the
12.5 dB attenuation limit, and attenuation greater than 12.5 dB for
only one atmosphere crossing is represented in gray on the Fig.
\ref{homorealistic}-(b)). Some radius/density combinations are then prejudicial 
to the Radar experiment and
its main objective of surface imaging. The model we considered here 
is very simple but shows that it is worthwhile to study the possible impact of atmospheric
particles (aerosols, hail particles, drizzle and rain droplets) on
the radar experiment. It allows also to define some first order
ranges of radius and density for which the radar could encounter
difficulties to see the surface of Titan. It would be the case
with, for example, a 500 km thick atmosphere constituted of
particles of $6\:\mu m$ radius with a $1000\:cm^{-3}$
density, which is not completely unexpected or unrealistic at
first order when we refer to distributions retrieved from
observations or microphysical calculations including rain
occurrence (particles are then concentrated in a thinner layer, but
are much bigger than ten microns \citep{1988toonb,1990frere}).
The scattering cross section increases with the sixth power of the
radius of the scatterers and constitutes the more sensitive parameter.
These first results lead us to the conclusion that Titan's atmosphere
might not be as transparent to Cassini radar frequencies as it was
expected. Radar pulses could at least be affected and distorted,
even if they encounter smaller particles than those we considered
in this section.

\subsubsection{Heterogeneous atmospheres and cloud layers}

In this section, we take into account as a modeling refinement the
possible heterogeneity of aerosol properties with altitude. We
do not treat the atmosphere as an ideal homogeneous layer anymore
but as a more complex medium as predicted by numerous published
models.

\begin{sidewaysfigure}
\begin{center}
\resizebox{\hsize}{!}{\includegraphics{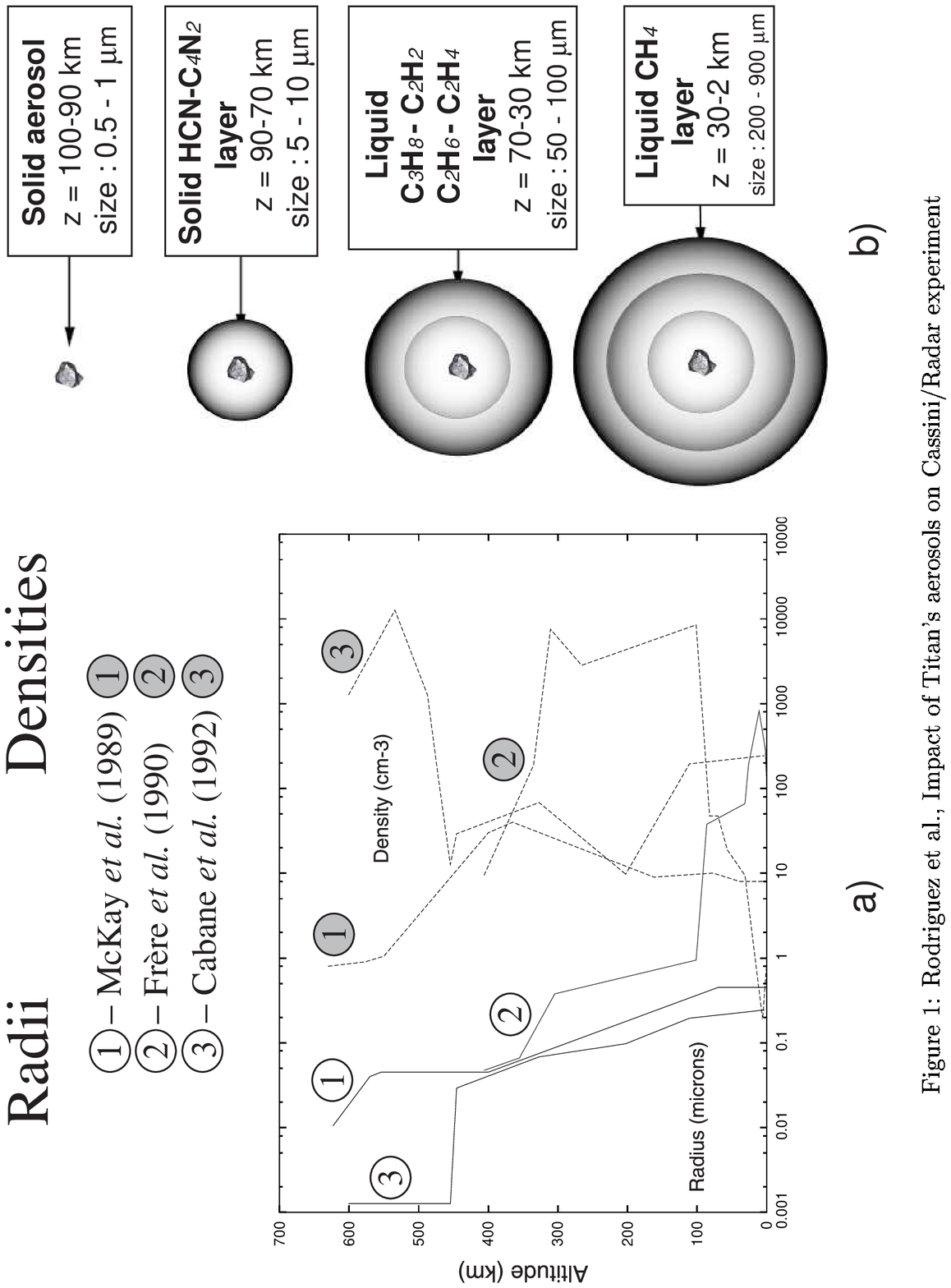}}
\caption{Rodriguez et al., Impact of Titan's aerosols on
Cassini/Radar experiment} 
\label{heterogmodels}
\end{center}
\end{sidewaysfigure}

\begin{sidewaysfigure}
\includegraphics[width=23cm]{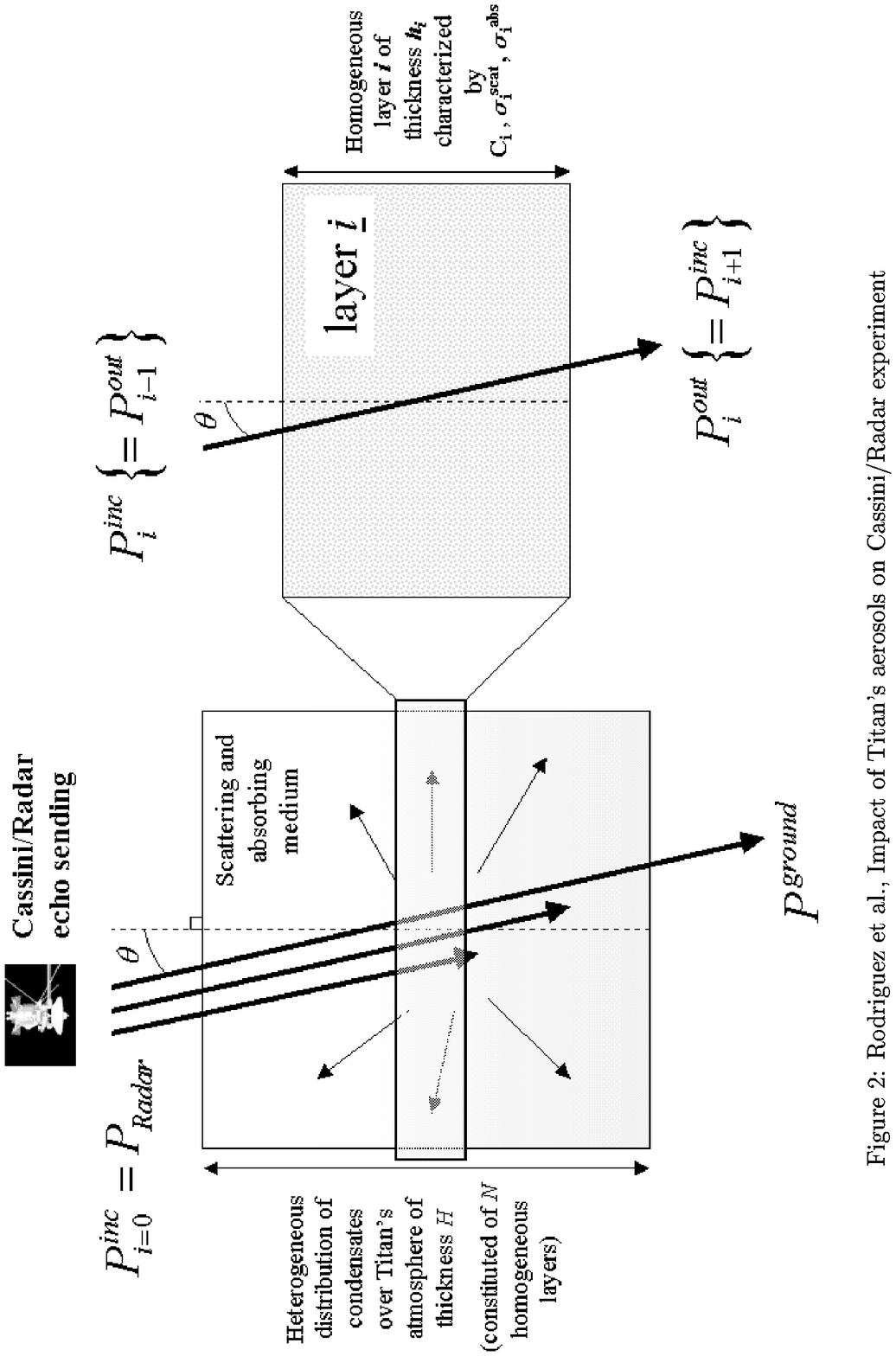}
\caption{Rodriguez et al., Impact of Titan's aerosols on
Cassini/Radar experiment} \label{model}
\end{sidewaysfigure}

\begin{figure}
\begin{center}
\resizebox{\hsize}{!}{\includegraphics{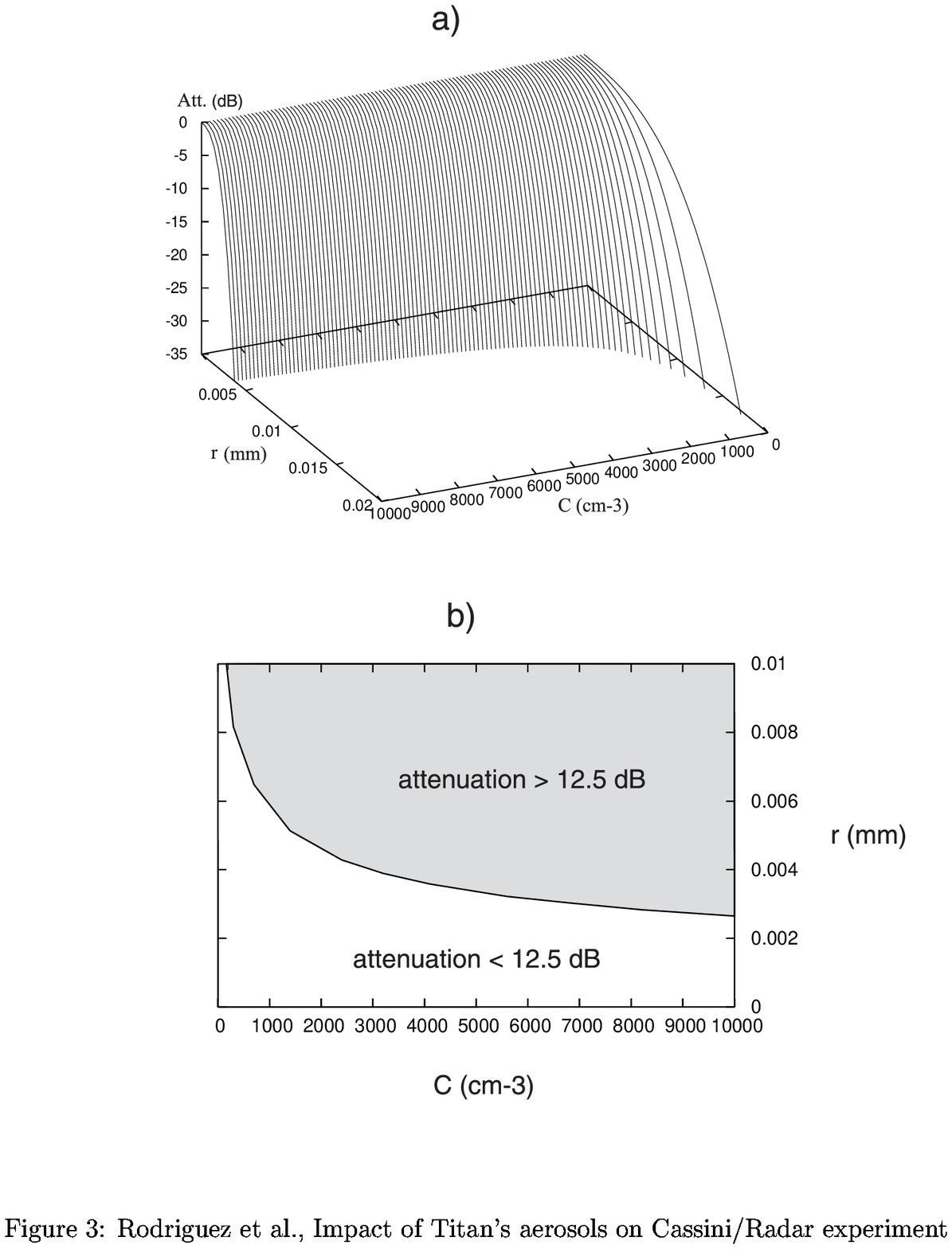}}
\caption{Rodriguez et al., Impact of Titan's aerosols on
Cassini/Radar experiment} \label{homorealistic}
\end{center}
\end{figure}

We used the heterogeneous distribution of particles discussed in
section \ref{distributions} in order to simulate a more realistic Titan's
atmosphere (see Fig.\ref{heterogmodels}-(a)).
We also included as a reference two semi-homogeneous models for Titan's 
atmosphere inferred from Pioneer and Voyager image analysis 
\citep{1981smith,1982smith,1982tomasko}. High-resolution views of 
the limb at high-phase angles revealed the presence of several layers of 
aerosols: an optically thin layer of haze 
$\sim\:50\:km$ thick roughly $100\:km$ above a main aerosol layer. The 
aerosol particle average size in the main haze layer was found to be
between $0.1$ and $0.5\:\mu m$ for a real part of refractive index
$n_r$ chosen between $1.7$ and $2$ at visible wavelengths. For the
upper haze layer, a mean value for the particle radius was taken to
$0.3\:\mu m$. Particle densities were calculated by
\citet{1980toon}. They found $10^4\:cm^{-3}$ for the first tenth
of the visible optical depth in Titan's atmosphere and
$10^3\:cm^{-3}$ at an optical depth of 1. In addition to the
detached haze layer, and above it, \citet{1983rages&pollack}
discovered a small but persistent feature in the Voyager 2
extinction profiles, characteristic of another haze layer with the
following physical properties: roughly 50 km thick, composed of
$3\:\mu m$-particles with a density of $\sim\:0.2\:cm^{-3}$ (cf. Table 5). 
We considered a mean aerosol permittivity $\varepsilon_r =
2.2 - j0.05$. Errorbars on attenuation come from uncertainties on tholin 
permittivity (the minimum value for attenuation was
calculated with $\varepsilon^{'} = 2.1$ and $\varepsilon^{''} =
0$, and the value maximum with $\varepsilon^{'} = 2.1$ and
$\varepsilon^{''} = 0.1$). 

We considered also the probable occurrence of hydrocarbon rain at low
altitudes. It was treated differently according to two scenarios: 
(1) We simulate the rain as another independent
layer in addition to aerosols (\citet{1988toonb} and \citet{1995courtin}
approach) and, (2) we consider rain as the result of continuous
growing of aerosols (\citet{1990frere} approach). In case (1), we
construct the cloud cover by substituting the \textquotedblleft
dry\textquotedblright $\:$ particle distribution (two homogeneous 
multi-layer models as defined earlier and 
models presented on Fig. \ref{heterogmodels}-(a)) by a
homogeneous layer containing methane droplets inspired by \citet{1988toonb}
 (cf. Table 1). The vertical extension of the cloud is
supposed to range between altitudes of 30 and 10 km. We fixed an
average methane droplet radius to $2\:mm$ and their permittivity to
$\varepsilon_r = 1.7 -j0.015$, based on the hypothesis that the
raindrops are only composed of pure liquid methane and that the tholin nuclei
can be neglected (the liquid phase hugely dominates the
composition of the droplet). The way we control the density of
the cloud was inspired by recent work of
\citet{1997samuelson&mayo} and \citet{1997guez}. They suggested
that liquid hydrocarbons may condense only from a limited subset of
the aerosols falling through the saturated layer, in particular
for kinetic reasons. \citet{1997guez} introduced
the concept of nucleation rate. Indeed, the condensation process
efficiency is highly dependent on the contact angle of condensed
materials on the nucleus. This is an essential parameter to estimate
the compatibility between nucleus and the condensed phase. As contact
angles are very unconstrained parameters, we treated the problem in a
simple way thanks to a condensation efficiency factor $f$ that can be free to vary between 0 and 100\%.
Using this coefficient, we simulate the fact that only a fraction of aerosols
can supply the cloud in droplets at the condensation level: for an atmosphere 
sampled every kilometer, cloud concentration is then given by $C_{droplets}
(z=30\:km) = f\times C_{aerosols}(z=31\:km)$. We checked that 
densities obtained for the top of the methane cloud ($z=30 km$) never exceed the limits 
imposed by \citet{1988toonb}, whatever higher aerosol distribution can be.
The f parameter was fixed equal to $10^{-4}$\%, giving 
particle concentrations at the cloud top of 1000, 1000, 8,
and $250\:m^{-3}$ respectively for the atmospheric models A, B, C and D (cf. Table 5). 
The cloud's particle radius and concentration are then
constant with respect to the altitude from the cloud top (30 km) to
its base (10 km). Aerosols which don't participate in the condensation
process are neglected, and Titan's
atmosphere is then only constituted of a homogeneous cloud of methane
rain droplets between 30 and 10 km. Below 10 km, as \citet{1993lorenza} showed,
pure methane droplets should evaporate and resupply aerosol layers,
forming a \textquotedblleft ghost\textquotedblright $\:$ mist of
aerosols close to the surface. In order to simulate rain sublimation in the
last kilometers of Titan's troposphere, we stop the rain layer
below 10 km and replace it by the \textquotedblleft
dry\textquotedblright $\:$ scenario we used to describe the
atmosphere above 30 km. 

In case (2), we simply used the
heterogeneous distribution computed by \citet{1990frere} (Fig.
\ref{heterogmodels}-(a)) as the complete description of the fate
of particles over the full extent of Titan's atmosphere, even
at low altitudes where condensation should occur. As they treated
the differential condensation of a great number of hydrocarbons
and nitriles, we chose an averaged permittivity for the raindrops
inferred from those measured for a homogeneously mixed hydrocarbon-nitrile liquid: 
$\varepsilon_r = 1.8 - j0.002$ (as for case
(1), electrical properties of the condensation nuclei were
neglected). In this case, the artificial factor $f$ has no sense
anymore. All aerosols participate in the condensation process.
The density and the size of the rain droplets below 100 km are
thus self-determined by the calculations of \citet{1990frere} thanks to
modeling of collisional and sticking processes.

All the models used in our calculations are summarized in Table
5.

{\bf [Table 5]}

{\bf [Figure 4]}

\begin{figure}
\begin{center}
\resizebox{\hsize}{!}{\includegraphics{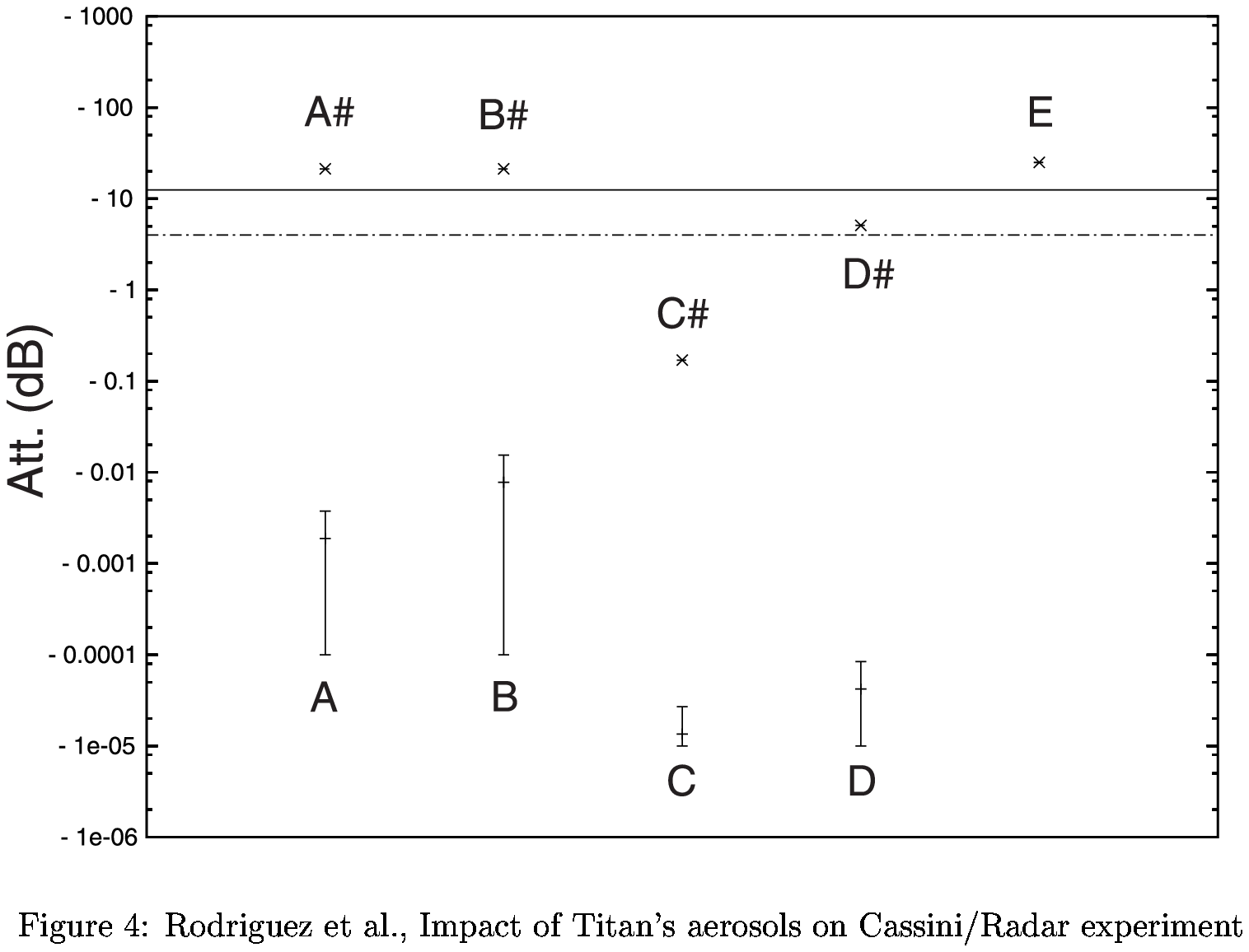}}
\caption{Rodriguez et al., Impact of Titan's aerosols on
Cassini/Radar experiment} \label{attweterrors}
\end{center}
\end{figure}

Figure \ref{attweterrors} shows the results obtained for the nine
models described in Table 5. Cases A, B, C and D correspond to attenuations
induced by a dry atmosphere, only constituted of aerosol hazes.
The attenuation never exceeds -0.02 dB, even for the least
favorable scenario B (containing the highest particle size and
density over the greatest part of the atmosphere), and are
globally negligible ($\leq\:-10^{-3}$ dB). We can conclude that,
without rain occurrence in the troposphere, Titan's atmosphere
should be totally transparent for the 13.78 GHz radar of the Cassini
orbiter. When we consider cases A$\sharp$, B$\sharp$, C$\sharp$, D$\sharp$ and E,
the conclusions are quite different. The addition of a rain layer
in the middle troposphere increases the signal attenuation by at
least four orders of magnitude. For D$\sharp$, A$\sharp$-B$\sharp$ (rain
attenuation overflows the attenuation induced by aerosols and the
difference between scenario A$\sharp$ and B$\sharp$ is then small;
thus A$\sharp$ and B$\sharp$ can be considered as one unique
scenario), and E models, attenuation reaches
respectively -5 dB, -20 dB and -25 dB. On the other hand, the
value of attenuation is still low for scenario C$\sharp$ (-0.17
dB), due to the fact that the rain cloud modeled for this case presents
a very low density. This parameter is controlled by the factor
$f$, fixed to $10^{-4}$\%, and the density of aerosols just above
30 km. For scenario C$\sharp$, the cloud has a density of only
$8\:m^{-3}$, in comparison to models A$\sharp$ and D$\sharp$
which have a raincloud density of respectively $1000\:m^{-3}$
and $250\:m^{-3}$. We have to keep in mind that the titanian
aerosol modeling of \citet{1989mckay} (scenario C and C$\sharp$)
was rather simple, and the aerosol density it predicts around
30 km is certainly less accurate than the one obtained from \citet{1990frere}
(scenario E) and \citet{1992cabane} (scenario D and D$\sharp$). We
are convinced that, even if clouds in Titan's atmosphere are 
transient and rare phenomena, they could surely affect the
radar signal. For two of the scenario we studied (A$\sharp$ and E),
clouds should totally attenuate the radar pulse before it 
reaches Titan's surface. For model D$\sharp$, it should also cause
sufficient attenuation to possibly blur surface images by
diminishing the contrast between distinct surface echoes.

{\bf [Figure 5]}

\begin{figure}
\begin{center}
\includegraphics[height=23cm]{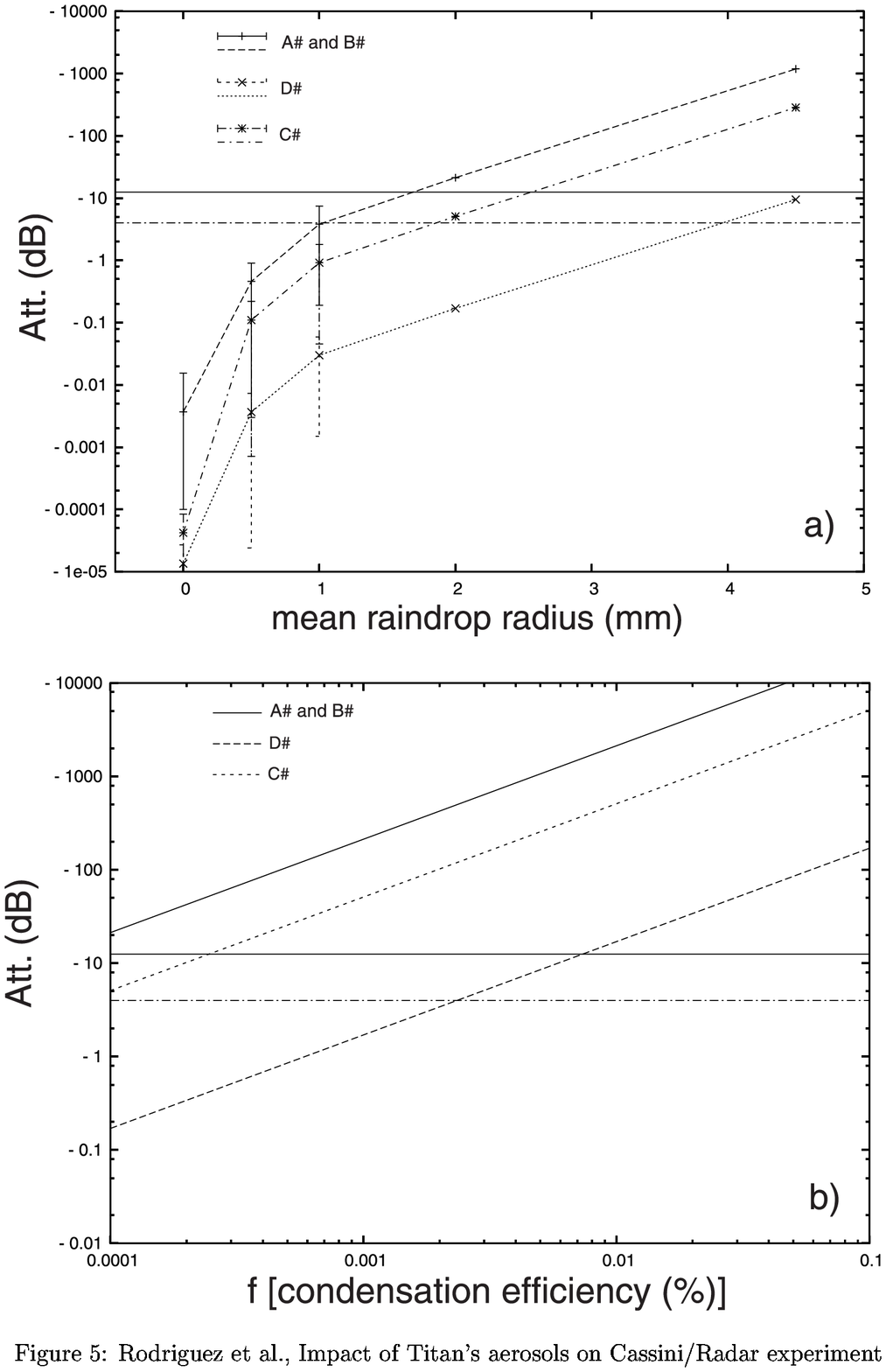}
\caption{Rodriguez et al., Impact of Titan's aerosols on
Cassini/Radar experiment} \label{attwetvsrf}
\end{center}
\end{figure}

The rain layer we used in addition to models A, B, C, and D in
order to simulate the condensation process in Titan's low atmosphere
is intrinsically artificial in its construction. We 
explored the way the attenuation may vary if we free some badly
constrained parameters of this layer: radius of droplets and
condensation efficiency f. The radius can vary between 0 mm
(no rain) and 4.5 mm (limit imposed by aerodynamics
\citep{1993lorenza}). For the parameter $f$, even if there are no
means at the present time to physically impose a constraint on it, we can consider that 
the value we chose ($f = 10^{-4}$\%, {\it i.e.} one aerosol 
per million acts as a condensation nucleus) in order to match the cloud
densities predicted by \citet{1988toonb} is very low. Higher values (up to 0.1\%) 
cannot be totally excluded and
should cause much more attenuation. We present in Fig.
\ref{attwetvsrf} the results of this parametric study: the attenuation as a
function of rain droplet radius (Fig. \ref{attwetvsrf}-(a)) and
condensation efficiency (Fig. \ref{attwetvsrf}-(b)) for models
A$\sharp$-B$\sharp$, C$\sharp$ and D$\sharp$. The increase of $r$
and $f$ induces a drastic rise of attenuation for all the
cases considered. Attenuations are much greater than unity as soon
as $r$ reaches 2 mm, and $f$ reaches 0.001\%. Figure \ref{attwetvsrf} demonstrates the
high sensitivity of the attenuation to $r$ and $f$, 
which are unfortunately weakly constrained.

\subsection{Surface reflectivity}

As aerosols are supposed to deposit on Titan's surface and form a
\textquotedblleft dirty ice\textquotedblright $\:$ mixture, we
tried to simulate the radar response of a surface covered with
tholins. It allows us to refine the first hypothesis we made in
the previous section of a perfectly reflecting surface ({\it i.e.} all the
energy reaching Titan's surface is backscattered to the instrument).

\begin{figure}
\begin{center}
\includegraphics{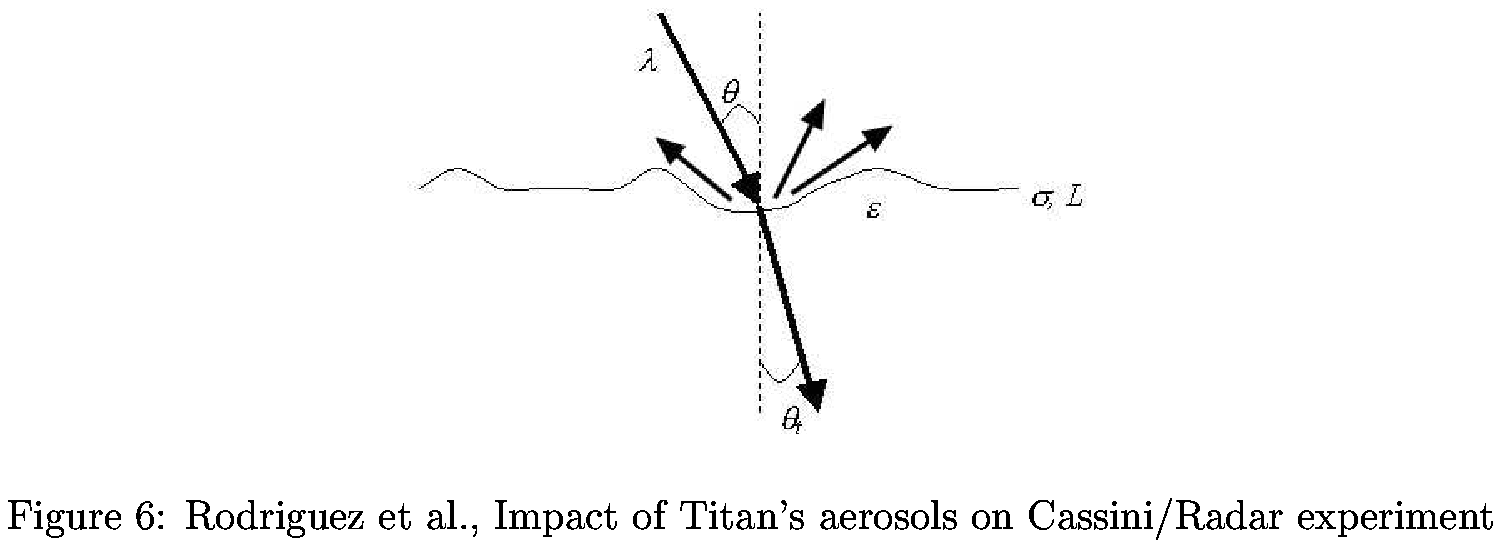}
\caption{Rodriguez et al., Impact of Titan's aerosols on
Cassini/Radar experiment} \label{geomsurf1}
\end{center}
\end{figure}

Since the Cassini Radar experiment will operate in the Ku-band
($f=13.8\:GHz$, $\lambda=2.17\:cm$, HH polarization
\citep{1991ieee}, most of Titan's surface is likely to appear as
rather rough for the radar \citep{1999dierking}. We shall then
consider surface scattering models developed for medium rough to
very rough surfaces, {\it i.e.} the Physical Optics (PO) model and the
Geometric Optics (GO) model, which were derived from the Kirchhoff
model under the scalar approximation and the stationary-phase
approximation respectively. These models are relevant to evaluate
performances of a side-looking imaging radar. We shall consider
here a single homogeneous layer covered by tholins (characterized
by its averaged dielectric constant $\varepsilon=2.2-0.05j$), water ice 
or liquid methane, and characterized by its surface
roughness. The latter is defined by the surface correlation
function $\rho(x)$ (a Gaussian function is assumed here), the
height standard deviation $\sigma$, and the correlation length $L$
(cf. Fig. \ref{geomsurf1}).

{\bf [Figure 6]}

The PO model validity range is defined by the following
conditions:

\begin{equation}
    \frac{\sqrt{2}\sigma}{L} < 0.25,\quad kL > 6\quad  and\quad  L^2 > 2.76\sigma\lambda
\end{equation}

\noindent{where $k$ is the wave number of the incident plane wave
($k=2\pi/\lambda$). The HH noncoherent scattering coefficient is
then given as a function of the incidence angle $\theta$ by
\citet{1982ulaby}:}

\begin{footnotesize}
\begin{equation}
\label{equasurf1}
    \sigma_{HH}^0(\theta)=2k^2\cos^2\theta\Gamma_H(\theta)e^{-(2k\sigma\cos\theta)^2}\cdot\sum_{n=1}^{\infty}(4k^2\sigma^2\cos^2\theta)^n/n!\:\int_0^{\infty}\rho^n(x)J_0(2kx\sin\theta)xdx
\end{equation}
\end{footnotesize}

\noindent{where $J_0$ is the zeroth-order Bessel function of the
first kind, $\Gamma_H(\theta)=\mid R_H(\theta)\mid ^2$ is the
Fresnel reflectivity, and $\rho(x)=\exp(-x^2/L^2)$ is the Gaussian
surface correlation function.}

The GO model validity range is expressed by:

\begin{equation}
    (2k\sigma\cos\theta)^2 > 10,\quad kL > 6\quad  and\quad  L^2 > 2.76\sigma\lambda
\end{equation}

\noindent{and the HH (or VV) noncoherent scattering coefficient is
given as a function of the angle of incidence $\theta$ by
\citet{1981fung}:}

\begin{equation}
\label{equasurf2}
    \sigma_{HH}^0(\theta)=\frac{\Gamma(0)e^{-\tan^2(\theta)/2m^2}}{2m^2\cos^4\theta}
\end{equation}

\noindent{where  $m=\frac{\sqrt{2}\sigma}{L}$ is the rms slope for
a Gaussian surface and $\Gamma(0)$ is the Fresnel reflectivity
evaluated at normal incidence:}

\begin{equation}
    \Gamma(0) = |\frac{1-\sqrt{\varepsilon/\mu}}{1+\sqrt{\varepsilon/\mu}}|^2
\end{equation}

\noindent{(we consider here a magnetic permeability
$\mu=1$).}

{\bf [Figure 7]}

\begin{figure}
\begin{center}
\includegraphics{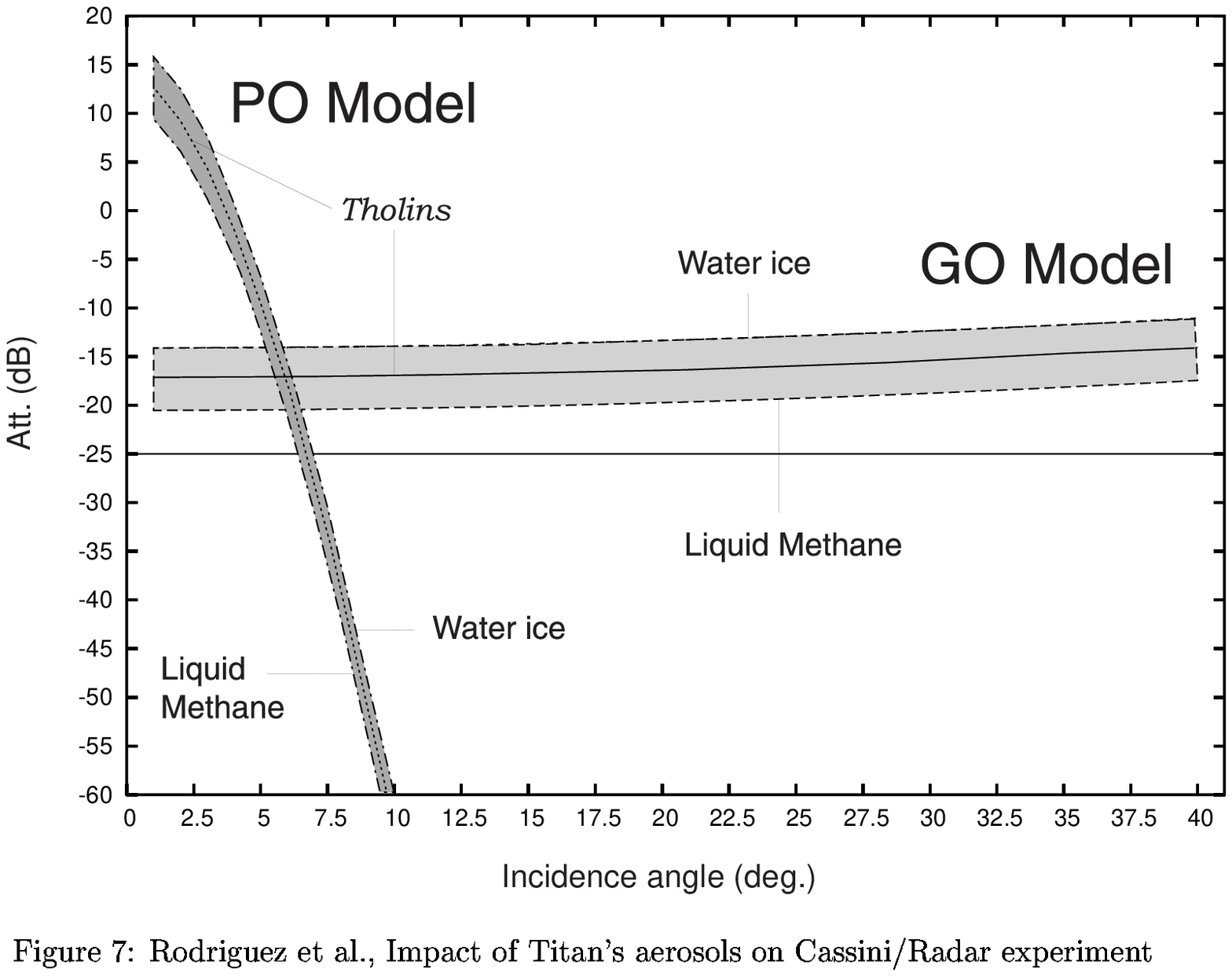}
\caption{Rodriguez et al., Impact of Titan's aerosols on
Cassini/Radar experiment} \label{powersurf2}
\end{center}
\end{figure}

Figure \ref{powersurf2} presents the radar backscattered power for
two types of surfaces: a relatively
smooth surface with $\sigma=0.5\:cm$ and $L=25\:cm$ described by
the PO model, and a rough surface with $\sigma=7\:cm$ and
$L=10\:cm$ described by the GO model. Calculations were 
performed for surfaces covered by tholins, water ices and liquid
methane. Uncertainties on experimental tholin dielectric constant 
were considered, but the resulting imprecision on backscattered power actually 
stays between the values obtained for water ice and liquid methane.
For the mean look angle foreseen for the imaging mode
of the Cassini radar ($\theta_L\sim11^{\tiny o}$
\citep{1991ieee}), we can see that the backscattered power for a
smooth flat surface is very low (about -80 dB), much lower than
the noise equivalent $\sigma_0$ of the instrument announced to be
around -25 dB \citep{1991ieee}. On the contrary, a rough and flat
surface should present a much higher return, about -17 dB. As a first approximation, 
the Cassini imaging radar
should only see very rough regions covered by tholins (land
surfaces ?) whereas smooth and flat surfaces 
(seas and lakes ?) should appear as dark zones. Nevertheless, for
non-flat smooth regions, Fig. \ref{powersurf2} shows that slopes
facing the radar with an incidence angle $\theta$ less than
$7^{\tiny o}$ would produce a backscattered power higher than -25
dB. Considering the mean look angle $\theta_L$, it corresponds to
a slope ranging between $4^{\tiny o}$ and $28^{\tiny o}$, which
represent a large fraction of slope values that can be observed on
planetary surfaces. Results obtained here should be of course 
considered as indicative, since very little is known about Titan's
surface composition and roughness. In particular, due to the low
loss tangent of tholins
($\tan\delta=\varepsilon^{''}/\varepsilon^{'}=0.02$) and other surface materials
such as water ice, the penetration depth $D_P$ of the incident Ku-band wave given by:

\begin{equation}
    D_P = \frac{\lambda}{4\pi}\{\frac{\mu
    \varepsilon^{'}}{2}[\sqrt{1+(\frac{\varepsilon^{''}}{\varepsilon^{'}})^2}-1]\}^{-1/2}
\end{equation}

is about 10 cm. This could give rise to some attenuation and
volume scattering effects that could change the radar
backscattered power computed from Eq. (\ref{equasurf1}) and
(\ref{equasurf2}). A mixture between silicate and tholins should
also lead to higher values of $\varepsilon$, and then to a higher reflectivity of the surface.

If we make the link between the results we obtained for
atmospheric attenuation and the surface backscattered power, it
leads to more realistic conclusions for the attenuating role of
the atmosphere. The surface should not be perfectly reflective for
the radar wave as we hypothesized in the previous section. In the
favorable case of a rough surface, the reflected
signal could still lose 17 dB, and the limit of the acceptable atmospheric
attenuation should then be reduced to -4 dB instead of -12.5 dB if 
considering a radar sensitivity limit of $-25\:dB$. This dramatically increases 
the impact of an attenuating atmosphere
on the Radar experiment (see Fig. \ref{attweterrors} and
\ref{attwetvsrf} with the attenuation limit fixed at -4 dB).

\section{Conclusion and perspectives}

The Cassini mission has been designed to send a
spacecraft to the planet Saturn, and deploy an instrument probe,
Huygens, that will descend to the surface of Saturn's moon Titan.
The radar instrument onboard the orbiter is expected to bring the
scientific community new information to characterize 
the surface of Titan. We are convinced that correct
interpretation of the measurements made with the
Cassini Radar instrument requires simulations and
laboratory measurements to anticipate possible atmospheric
effects and then better invert and interpret radar data.

In this paper we presented a study of the behaviour of a $13.78\:GHz$
plane wave going through Titan's atmosphere and
backscattered by its surface. Titan's atmosphere was modeled by a
diluted medium filled with spherical particles of size,
density and permittivity estimated from what could be found in the
literature and laboratory experiments. We considered
one homogeneous atmosphere model and five heterogeneous models including
or not condensation that could occur in the lowest part of
titanian atmosphere. In order to fix some parameters of our models, we 
conducted a series of experiments in order
to derive the dielectric properties around $13.78\:GHz$ of Titan's
aerosol analogs produced in laboratory.

The results we obtained in terms of radar power attenuation after
crossing the simulated Titan's atmosphere are of two kinds:
\begin{itemize}
\item the \textquotedblleft dry\textquotedblright $\:$ heterogeneous atmospheric
models, only taking into account the aerosol component, do not
cause any attenuation and the radar wave will reach the surface of
Titan without any losses. The particle radius never exceeds 
$1\:\mu m$ (far from the critical limit around $10\:\mu m$ we estimated from the ideal
homogeneous case) and the corresponding attenuation never exceeds $-0.01\:dB$. In such
cases, Titan's atmosphere is totally transparent to microwaves and its effect
should be totally neglected in the future processing of Cassini
radar data.
\item on the contrary, when we add in our simulation a rain layer in the last few
kilometers, the attenuation reaches and rapidly exceeds the
instrument sensitivity limit, due to the enhancement in size
of the particles encountered (that could reach a few millimeters).
A wave emerging from such a cloud layer could be so attenuated
that the radar antenna wouldn't be able to detect a returned echo
from Titan's surface. This could be a problem for the retrieval of Titan's
surface images, as well as for altimetry measurements, and it should be taken into account when
interpreting future data of the Cassini Radar experiment.
\end{itemize}
Simulation of the backscattered signal by Titan's surface covered
by tholins also shows that a rather low return could be expected, even for
rough surfaces. This could dramatically increase the possible attenuation
role of Titan's atmosphere.

There is also a non-negligible risk for the
Cassini radar instrument to detect a false signal coming from a
reflective atmospheric layer, screening and
flooding the surface signal,
that could cause false interpretation of \textquotedblleft
surface\textquotedblright $\:$images. As the Cassini radar is a
multipurpose instrument, a way to discriminate between possible
atmospheric and surface echoes would be to combine altimetric and SAR data. It could then be
possible to avoid false interpretation of \textquotedblleft
non-surfacic\textquotedblright $\:$signal, if both altimetric and SAR acquisition 
can be performed in the same short time period. The altimetry
pulse shape could also be analyzed in order to detect an
atmospheric effect and get
unexpected information about Titan's lower atmosphere (rain
occurrence and extension of cloud systems, size, density and
velocity of the particles). In a flyby strategy which would
maximize the coverage of Titan's surface, the radar will
certainly not map the same region twice. In this case, using the
altimetry data would be the only way to discriminate possible
cloud layers from surface features. On the other hand, two
flybys over the same area could give the opportunity to
distinguish persistent features related to the surface from transient
ones, which are likely to be due to atmospheric phenomena.

The new analysis scheme we propose for the Cassini radar
measurements will be subject to further investigations. For this
purpose, we shall improve our present simulation of the
Cassini Radar instrument in order to model the behaviour of
the pulse it will send towards Titan's surface (not only power
attenuation, but also shape distortion). We shall also take
into account the doppler shift in frequency of the signal sent and
received by the orbiter, as could be produced by the relative motion of the atmospheric
scatterers. In order to properly prepare for the analysis of future Cassini
radar data, we also need for more accurate measurements of
haze and rain properties (size, density and dielectric
constant) as we showed the sensitivity of our calculations
to these parameters. Rain droplet radius and
condensation efficiency in the low atmosphere are in particular crucial parameters. Such information
could also be retrieved from the combined interpretation of
independent observations of Titan's atmosphere coming from the other
instruments onboard the Cassini orbiter or Huygens probe, as well as
results of future rain modeling.

\noindent{{\bf ACKNOWLEDGMENTS}}

We sincerely thank PIOM's director, Jean-Paul Parneix, and LISA's
director, Fran\c{c}ois Raulin, who welcomed us in their
laboratories and gave us the opportunity to instructively
collaborate with some of their team members. We are also very grateful to Jonathan Braine 
who patiently checked the English language of the paper. This work was
supported by the French {\it Programme National de Plan\'etologie}
of the Institut National des Sciences de l'Univers, CNRS.

\bibliographystyle{plainnat}


\clearpage

\pagestyle{empty}

\begin{table}[htbp]
    \caption{Methane rain properties retrieved from modeling by Toon {\it et al.} (1988) and Courtin {\it et al.} 
    (1995).}
        \label{raintoon}
    \[
        \begin{array}{p{0.5\linewidth}p{0.5\linewidth}}
    \hline
    \noalign{\smallskip}
Physical property & Value  \\
    \noalign{\smallskip}
    \hline
    \noalign{\smallskip}
Cloud extension & $20\:km$ \\
Altitude of cloud top & $z=30-40\:km$ \\
Mean raindrop radius & between $50\:\mu m$ and $3\:mm$ (Toon only)\\
Raindrop density & between $1$ and $3500\:m^{-3}$ (Toon only)\\
Optical thickness & $\leq2$ (far-infrared) \\
    \noalign{\smallskip}
    \hline
        \end{array}
    \]
\end{table}

\clearpage

\begin{sidewaystable}[htbp]
    \caption{Set of samples supplied by simulations at LISA}
        \label{samples}
    \[
        \begin{array}{p{0.28\linewidth}p{0.39\linewidth}p{0.30\linewidth}}
    \hline
    \noalign{\smallskip}
        & Synthesis conditions   & Volume of tholins deposited \\
    \noalign{\smallskip}
    \hline
    \noalign{\smallskip}
Set n$^{\tiny o}$1   & $V=2.2\:kV$ &  \\
(3 samples on glass tablets) & $I=80\:mA$ &  \\
 & gas mixture of $N_2/CH_4$ (in proportion $98:2$) & \\
 & $p=3\:mbar$ & \\
 & deposit duration : 4h25 & $<\:1.5\: mm^3$ \\
 & & \\
 & $V=2.6\:kV$ &  \\
 & $I=50\:mA$ &  \\
 & gas mixture of$N_2/CH_4$ (in proportion $98:2$) &  \\
 & $p=4\:mbar$ & \\
 & deposit duration : 5h50 & $\sim\:1.5\: mm^3$ \\
 & & \\
 & $V=2.6\:kV$ &  \\
 & $I=50\:mA$ &  \\
 & gas mixture of $N_2/CH_4$ (in proportion $98:2$) &  \\
 & $p=4\:mbar$ & \\
 & deposit duration : 17h00 & $\sim\:8\: mm^3$ \\
 & & \\
Set n$^{\tiny o}$2  &  $V=2.8\:kV$  &  \\
(1 glass tablet + 1 silica cylinder) & $I=40\:mA$  &   \\
 & gas mixture of $N_2/CH_4$ (in proportion $98:2$) & \\
 & $p=4\:mbar$ & \\
 & deposit duration : 6h45 & $<\:8\: mm^3$ \\
 & deposit duration : 9h40 & $\sim\:16\: mm^3$ \\
 & & \\
Set n$^{\tiny o}$3  & $V=2.5\:kV$   &  \\
(1 silica cylinder) & $I=30.5\:mA$  & \\
 & gas mixture of $N_2/CH_4$ (in proportion $98:2$)  & \\
 & $p=1\:mbar$ & \\
 & deposit duration : 44h20 & $\sim\:2\: mm^3$ \\
    \noalign{\smallskip}
    \hline
        \end{array}
    \]
\end{sidewaystable}

\clearpage

\begin{table}[htbp]
    \caption{Tholin dielectric constants measured in resonant cavities.}
        \label{diel}
    \[
        \begin{array}{p{0.22\linewidth}p{0.33\linewidth}p{0.39\linewidth}}
    \hline
    \noalign{\smallskip}
        & Cavity used & Dielectric constant \\
    \noalign{\smallskip}
    \hline
    \noalign{\smallskip}
Set n$^{\tiny o}$1  & 2.45 GHz &  $\varepsilon_r = 2.26 - j0.12$ \\
 & & \\
Set n$^{\tiny o}$2  & 2.45 and 10 GHz & $\varepsilon_r = 2.2 - j0.002$ \\
 & & \\
Set n$^{\tiny o}$3  & 10 GHz & $\varepsilon_r = 2.03 - j0.022$ \\
    \noalign{\smallskip}
    \hline
        \end{array}
    \]
\end{table}

\clearpage

\begin{sidewaystable}
\begin{footnotesize}
    \caption{Microwave dielectric constants of some major materials 
    suspected to compose Titan's atmosphere and surface.}
        \label{sumperm}
    \[
        \begin{array}{p{0.25\linewidth}p{0.25\linewidth}p{0.25\linewidth}p{0.25\linewidth}}
    \hline
    \noalign{\smallskip}
Species & Dielectric constant & Frequency range of validity & Source \\
    \noalign{\smallskip}
    \hline
    \noalign{\smallskip}
Pure liquid $CH_4$ (94 K)& $\varepsilon_r=1.7 - j1.5\times10^{-2}$ & hundreds of MHz & Thompson and Squyres (1990) \\
 & & & \\
Perfectly mixed &  &  & \\
$CH_4$-$C_2H_6$ liquid & $\varepsilon_r=1.8 - j2\times10^{-3}$ & GHz & Singh (1979) \\
(around 100 K) & & & Sen {\it et al.} (1992) \\
 & & & \\
Pure $H_2O$ ice & $\varepsilon^{'}\:\sim\:3.1$  & $\sim\:$10 GHz & Ulaby {\it et al.} (1982)  \\
(between 90 and 273 K) & and very low loss tangent &  &  \\
 & & & \\
Pure $CH_4$-$C_2H_6$ & $\varepsilon^{'}\:\sim\:2-2.4$  & $\sim\:$10 GHz & Thompson and Squyres (1990) \\
ice (around 90 K) & and very low loss tangent &  & \\

    \noalign{\smallskip}
    \hline
        \end{array}
    \]
\end{footnotesize}
\end{sidewaystable}

\clearpage

\begin{sidewaystable}
\begin{footnotesize}
    \caption{Summary of the atmospheric models for Titan used in our
    simulations.}
        \label{scenar}
    \[
        \begin{array}{p{0.3\linewidth}p{0.5\linewidth}p{0.3\linewidth}}
    \hline
    \noalign{\smallskip}
Scenario label & Description & Physical properties \\
    \noalign{\smallskip}
    \hline
    \noalign{\smallskip}
 & & \\
Dry atmospheres & only aerosols & \\
 & & \\
A  & Homogeneous multi-layer model & Tholin dielectric constant\\
   & (with $0.1\:\mu m$-radius particles in the main haze layer) & $\varepsilon^{'} = 2.2\pm0.1$ \\
   & & $\varepsilon^{''} = 0.05\pm0.05$ \\
   & & \\
B  & Homogeneous multi-layer model & {\it idem} \\
   & (with $0.5\:\mu m$-radius particles in the main haze layer) & \\
   & & \\
   & & \\
C  & Inhomogeneous model taken from McKay {\it et al.} (1989) & {\it idem} \\
   & (see Fig. \ref{heterogmodels}-(a)) & \\
   & & \\
D  & Inhomogeneous model taken from Cabane {\it et al.} (1992) & {\it idem} \\
   & (see Fig. \ref{heterogmodels}-(a)) & \\
   & & \\
Wet atmospheres & aerosols + condensation droplets & \\
   & & \\
E  & Inhomogeneous model taken from Fr\`ere {\it et al.} (1990) & Droplet dielectric constant \\
   & including condensation at low alitude (below 100 km) & (mixed hydrocarbons)\\
   & (see Fig. \ref{heterogmodels}-(a)) & $\varepsilon_r=1.8 - j2\times10^{-3}$ \\
   & & \\
A$\sharp$ & Homogeneous multi-layers model A & Pure $CH_4$ rain layer \\
   & + methane rain layer between 30 and 10 km altitude & (nominal case) \\
   & & extension: 10-30 km \\
   & & radius: 2 mm \\
   & & $f = 10^{-4}$\% \\
   & & $\varepsilon_r=1.7 - j1.5\times10^{-2}$ \\
   & & \\
B$\sharp$ & Homogeneous multi-layers model B &  {\it idem} \\
   & + methane rain layer between 30 and 10 km altitude & \\
   & & \\
C$\sharp$ & Inhomogeneous model C & {\it idem} \\
   & + methane rain layer between 30 and 10 km altitude & \\
   & & \\
D$\sharp$ & Inhomogeneous D & {\it idem} \\
   & + methane rain layer between 30 and 10 km altitude & \\
    \noalign{\smallskip}
    \hline
        \end{array}
    \]
\end{footnotesize}
\end{sidewaystable}


\clearpage

\noindent{{\bf FIGURE CAPTIONS}}

\vspace{1cm}

\noindent{Figure 1. (a) Inhomogeneous models of Titan's aerosol
distribution inferred from microphysical calculations : (1) McKay 
{\it et al.} (1989), (2) Fr\`ere {\it et al.} (1990), and (3) Cabane {\it et
al.} (1992). (b) Schematic representation of differential
condensation processes occurring in the 100 last kilometers of 
Titan's atmosphere as modeled by Fr\`ere {\it et al.}
(1990). Differential condensation of hydrocarbons and nitriles
control the size, density and chemical composition of the aerosol
below 100 km. Aerosols are then of increasing size with decreasing
altitude, reaching a maximum radius of $\sim\:900\:\mu m$ around
10 km altitude. Below this altitude, sublimation of liquids
surrounding the aerosol leads to a reduction of the particle
size until hypothetical complete sublimation occurs 2-3 km
above the ground.}

\vspace{1cm}

\noindent{Figure 2. Geometry and description of the multi-layer model we used
for atmospheric transmission computation.}

\vspace{1cm}

\noindent{Figure 3. (a): Attenuation versus particle radius and
density for a 500 km thick homogeneous atmosphere with $\varepsilon_r=2.2-j0.05$;
(b): 12.5 dB-threshold versus radius and density of dielectric
spheres ($\varepsilon_r=2.2-j0.05$) for a 500 km thick homogeneous
medium. The grayed region represents atmosphere models
for which the radar is unable to see the surface
of Titan.}

\vspace{1cm}

\noindent{Figure 4. Attenuation caused by heterogeneous models
(aerosol only and aerosol+condensation droplets). The description and
properties of each scenario symbolized by letters in this figure
can be found in Table 5. The minimum values for attenuation were
calculated with $\varepsilon^{'} = 2.1$ and $\varepsilon^{''} =
0$, and the maximum values with $\varepsilon^{'} = 2.1$ and
$\varepsilon^{''} = 0.1$. The mean attenuation was calculated with a value of
$\varepsilon_r=2.2-j0.05$ contained within the
two attenuation extrema symbolized by the errorbars. The
comparison between \textquotedblleft dry\textquotedblright $\:$
models and \textquotedblleft wet\textquotedblright$\:$ ones
emphasizes the drastic increase in
attenuation due to rain occurrence. Errorbars are too small to be
seen in the \textquotedblleft wet\textquotedblright$\:$ model
case, due to the fact that aerosol attenuation - the only source
of incertitude here- become negligible compared to the rain
effect. The solid horizontal line shows the sensitivity threshold at
-12.5 dB , and the dotted line represents the -4 dB threshold.}

\vspace{1cm}

\noindent{Figure 5. (a): Attenuation for heterogeneous
models {\bf A$\sharp$}, {\bf B$\sharp$}, {\bf C$\sharp$}, and
{\bf D$\sharp$} with increasing droplet radius from no rain (0
mm) to the limit of 4.5 mm imposed by Lorenz (1993); (b):
Attenuation for models {\bf A$\sharp$}, {\bf B$\sharp$},
{\bf C$\sharp$}, and {\bf D$\sharp$} with increasing condensation
efficiency on aerosols (factor $f$) varying between
our nominal case ($10^{-4}\%$), in agreement with Toon {\it et al.} (1988),
 and $0.1\%$.
Errorbars on attenuation values due to uncertainties in the dielectric constant of tholin 
become insignificant beyond a
certain value of $r$ (2 mm) and are totally hidden by
attenuation due to rain as soon as $f$ is higher than $10^{-4}\%$. For (a) and (b), the 
dielectric constant of tholins was taken from the value we experimentally
measured: $\varepsilon_r = 2.2\pm0.1 -j0.05\pm0.05$. The solid
horizontal line shows the -12.5 dB sensitivity threshold and the dotted line the -4 dB threshold.}

\vspace{1cm}

\noindent{Figure 6. Geometry of the single layer scattering
problem.}

\vspace{1cm}

\noindent{Figure 7. Compared radar backscattered power for a
relatively smooth surface ($\sigma=0.5\:cm$ and $L=25\:cm$, PO
model) and a rough surface ($\sigma=7\:cm$ and $L=10\:cm$, GO
model) covered by tholins ($\varepsilon_r=2.2-j0.05$), water ice
($\varepsilon_r=3.1-j0.001$) or liquid methane ($\varepsilon_r=1.7-j0.015$).
The solid horizontal line shows the sensitivity threshold at -25 dB.}


\clearpage

\clearpage

\clearpage

\clearpage

\clearpage

\clearpage

\clearpage

\end{document}